# EI Videos

Michael Courtney, PhD and Lt. Col. (Ret.) Tom Slusher
United States Air Force Academy
Michael.Courtney@usafa.edu

Amy Courtney, PhD
BTG Research, PO Box 62541, Colorado Springs, CO 80962

**Abstract:** The Quantitative Reasoning Center (QRC) at USAFA has the institution's primary responsibility for offering after hours extra instruction (EI) in core technical disciplines (mathematics, chemistry, physics, and engineering mechanics). Demand has been tremendous, totaling over 3600 evening EI sessions in the Fall of 2010. Meeting this demand with only four (now five) full time faculty has been challenging. EI Videos have been produced to help serve cadets in need of well-modeled solutions to homework-type problems. These videos have been warmly received, being viewed over 14,000 times in Fall 2010 and probably contributing to a significant increase in the first attempt success rate on the Algebra Fundamental Skills Exam in Calculus 1. EI Video production is being extended to better support Calculus 2, Calculus 3, and Physics 1.

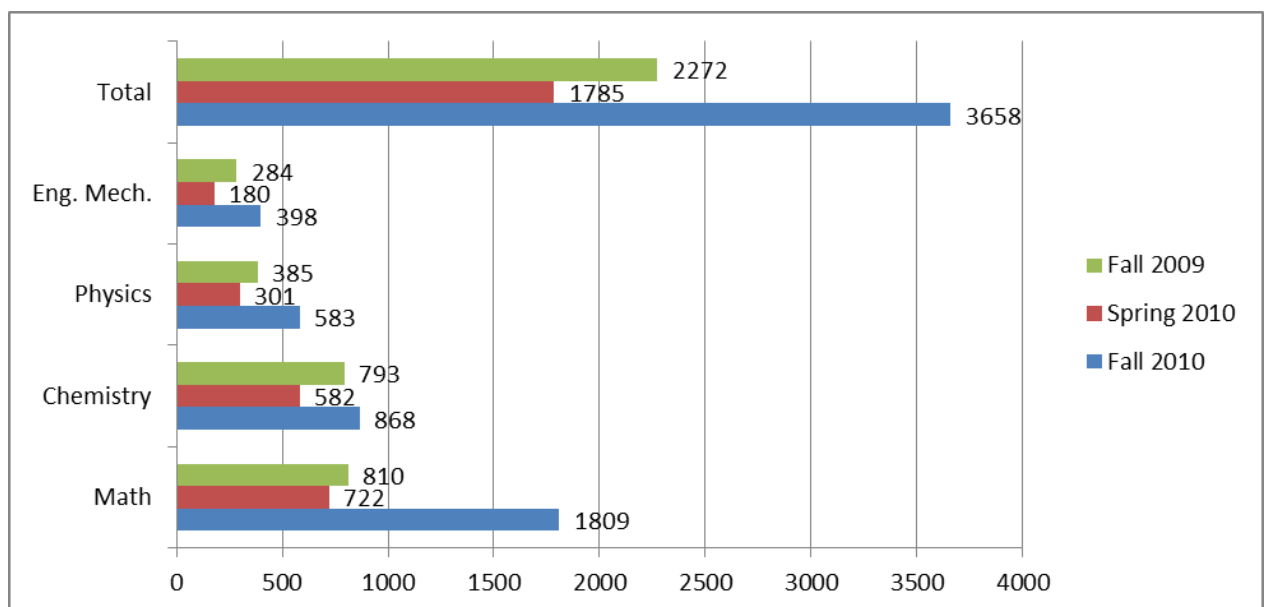

*Figure 1: USAFA evening EI sessions (in person) by subject, Fall 2009-Fall 2010, EI Video support was added for every lesson in Fall 2010.*

**Introduction**
Soon after USAFA began offering extra instruction (EI) services in core technical courses during evening hours (1800-2200), the demand and rapid growth revealed the need for force multipliers given the availability of only four (now five) full time faculty.[1] Figure 1 shows the number of evening EI sessions in each subject for the first three semesters of evening EI availability. In the Fall of 2010, demand for evening EI grew to 3600 evening EI sessions in the core classes, over 1800 sessions in the supported Math courses, including Pre-calculus, Calculus 1, Calculus 2, and Calculus 3, The Quantitative Reasoning Center (QRC) faculty quickly noticed that most EI needs fall into two phases: 1) modeling problem solving methods at cadet request and 2) coaching cadets as they work homework problems at the board. Because evening EI services are utilized disproportionately by cadets with weaker math backgrounds, it was observed that these cadets often need to see one or two worked examples (model phase) before the EI process can productively move to the coach phase. Modeling the same subset of assigned practice/homework problems repeatedly for different cadets quickly gave rise to the idea that a lot of the modeling phase could be accomplished by producing videos that can be posted on the USAFA intranet for cadets to view at their convenience, thereby leaving more EI face time for the coach phase of the EI process.

In Spring 2011, the USAFA QRC is working to build the available EI video library to at least one video pertaining to each lesson's assignment in Calculus 1, Calculus 2, Calculus 3, and Physics 1. We are in the process of prioritizing development for future semesters to serve the core Chemistry sequence and Physics 2. Manpower limits the ability to provide immediate support to a wider variety of problems or courses, since each 10 minute video requires between 1 and 2 hours to produce once the instructor is experienced in the process. Most instructors find themselves unsatisfied with their first few efforts and end up re-recording their first few videos.

**EI Video Pedagogical Philosophy and Method**
USAFA's academic departments do an excellent job presenting core technical material to cadets with average and above backgrounds in science and mathematics, and the institution is highly regarded for instructor availability for EI to meet the needs of cadets for whom the classroom presentation is not sufficient.[2] Evening EI and the EI Video programs are an extension of this availability. Figures 2 and 3 show that evening EI services are used disproportionately by groups traditionally underrepresented in the STEM disciplines and by cadets with below average academic backgrounds. (The academic composite is a measure of strength of a cadet's academic background and is intended to be a predictor of cadet GPA. The mean academic composite is between 3100 and 3200 for most incoming cadet classes.)

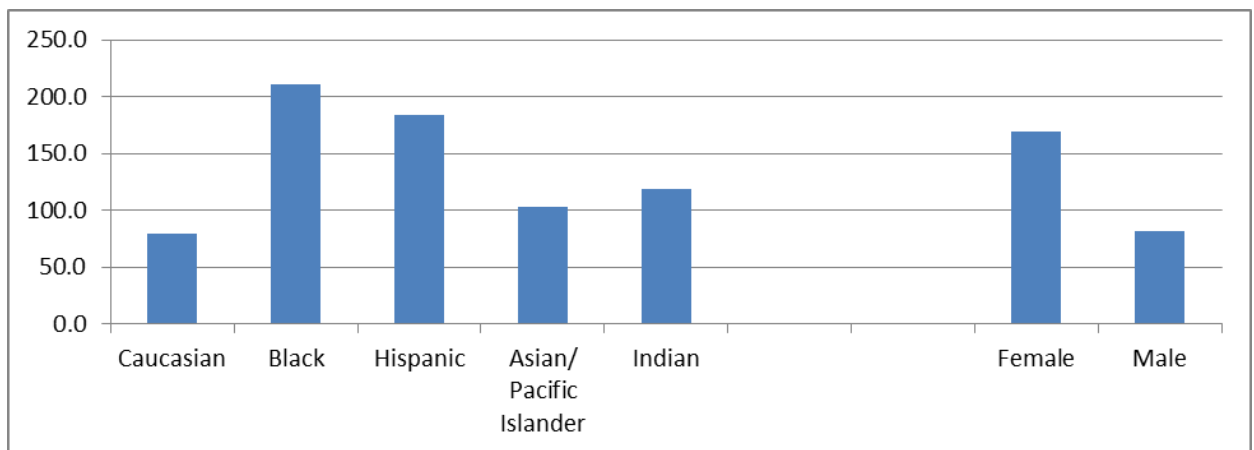

*Figure 2: Evening EI usage as percent of representation at USAFA. For example, since female cadets represent 35.3% of evening EI visits and 20.9% of underclassmen, the number reported above is 35.3%/20.9% (multiplied by 100%) = 169%. Thus numbers above 100 represent disproportionately high usage by the group. Numbers below 100 represent disproportionately low usage by the group.*

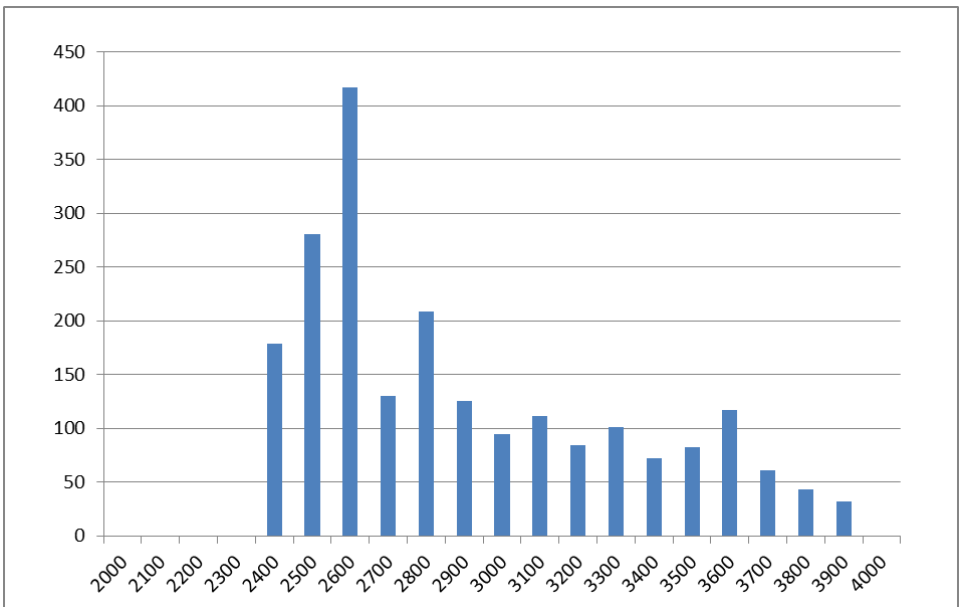

*Figure 3: Evening EI usage as percent of representation at USAFA for academic composites 2000-4000. For example, since cadets with academic composites from 3800 to 3899 represent 4.00% of underclassman but only 1.75% of evening EI visits, the number reported above is 1.75%/4.00% (multiplied by 100%) =43.7%. Thus numbers above 100 represent disproportionately high usage by the group. Numbers below 100 represent disproportionately low usage by the group.*

Recognizing that EI services are disproportionately utilized by students with weaker academic backgrounds, the pedagogy of the EI videos is aimed at the lowest quartile of students. This is a bit different from typical classroom instruction, which often tends to be aimed more at the middle. An example that might take only 5 minutes to discuss in class is slowed down to 10 minutes, and the video instructor is more intentional in most videos about developing a plan and working intermediate steps in greater detail. Experience has shown that cadets who understand the typical classroom level of presentation are not the most likely to be seeking EI. Furthermore, it is easy enough for stronger students to fast forward material that is overly detailed, but cadets who need more detail simply cannot slow things down enough to see things that are not on the video, and they can't raise their hands and ask questions if the instructor on the video skips a step they need to see.

Inspiration for the EI videos was drawn from many sources, but with particular emphasis on the MIT Video Lecture portion of their open course ware (http://ocw.mit.edu ) and on the MathTV channel at YouTube (http://www.youtube.com/user/MathTV ). Simply directing cadet traffic to these and other available high-quality internet video resources was considered. However, in the time squeeze that is cadet life, cadets were not interested in a search for the video that demonstrates the topic at hand in just the right way every time they have a question. In contrast, when course directors provide links in the syllabus directly to close analogues to assigned homework problems, cadets are quick to access the available resource since they are sure it directly pertains to the task.

Additionally, instructors both teach what we know and impart who we are; therefore, we also preferred an approach to EI Videos that would promote officer development as well as academic success. Since the goal is to develop character and not just teach academics, most EI Videos include a brief introductory vignette, a 30-120 second segment before the pedagogical portion encouraging cadets to form better habits, pointing out the military or practical applications of the topics being discussed, and/or sharing personal experiences related to the topic or the training process. One of the cadets' favorite vignettes is an instructor attempting dance moves from "Saturday Night Fever" and then admitting, "When I try to dance, I look like a dufus – because I haven't practiced." The vignette closes and transitions into the example problem with the admonition that without practicing the homework problems, watching EI Videos won't make them any better at math than watching "Dancing with the Stars" will make them a better dancer. Other vignettes feature an instructor with a barbell encouraging, "The math class is the weight room for the mind . . ." warning against "Five frequently fatal freshmen physics fantasies" [3] or holding a precision rifle and explaining the importance of mathematics in the profession of arms which is about "putting projectiles on target." Some example videos have been uploaded to YouTube, because a lot about the vignettes is hard to explain in writing, but easy to perceive. [4][5][6]

Detailed production tips are described in the appendix. None of the video instructors have been terribly excited about how they look and sound on video. The camera adds 20 pounds and seems to magnify every wrinkle and mannerism, every "um", "er", and pregnant pause while one collects a thought and considers the next phrase. Confidence and improved ability come with practice. We have learned to get over our vanities and get the job done putting well considered solutions on video. Teaching on video is a great tool for breaking bad habits and smoothing one's presentation. One video instructor lost 30 pounds to better present a good example of lifelong fitness on camera and in the classroom. The path to growth is jumping in and trying it.

**Results**

The EI Videos have been very well received by cadets, administration, and faculty alike. In Fall 2010, EI Videos were viewed over 14,000 times, which compares favorably with the 1800+ in person EI visits in Mathematics. It seems self-evident that students who avail themselves of EI opportunities will perform better. Our analysis of Fall 2009 data showed Calculus 1 students who visited in person for evening EI prior to the Algebra Fundamental Skills Exam had a first attempt score average (76%) nearly as high as the course wide average (78%), even though they scored much lower (52%) than the course wide average (60%) on the Algebra portion of the placement exam.

Since we have no way to track which cadets are viewing EI Videos, a direct comparison with in person EI is not possible. However, the course wide first attempt pass rate of the Algebra Fundamental Skills Exam in Math 141 increased from 53% in Fall 2009 (before EI videos) to 63% in Fall 2010 (after EI videos).[7][1] This increase could be due to other factors, such as significant changes and an emphasis in practicing for this event in USAFA's First Year Experience (FYE) course. However, many cadets directly attribute their success on the FSE to the availability of EI videos, and they overflow with gratitude and appreciation when they meet the video instructor in person.

One might wonder whether the availability of video EI is robbing cadets of the character development and obviously superior instruction of face time with a real instructor. The data suggest otherwise. In Fall 2010, only Mathematics courses were supported with EI videos, and growth of in-person evening EI in Mathematics far outpaced other disciplines. The Department of Mathematical Sciences does not track in-person daytime EI for the entire Calculus 1 course. However, the course director and instructors have offered anecdotes (and some sign-in logs) suggesting that daytime EI for the course was also increased significantly in Fall 2010 compared with Fall 2009. So it seems that EI videos have the potential to both meet demand for the modeling aspects of EI and create increased demand for the coaching aspects. The demographics of the incoming cadets in 2009 and 2010 were relatively constant (within 2%) with respect to incoming standardized test scores and the percentage of groups that are commonly underrepresented in STEM majors, so the growth of in-person EI is not attributable to a dramatic change in the incoming class. Video EI is also useful for students who miss class due to illness, travelling, or intercollegiate athletics.

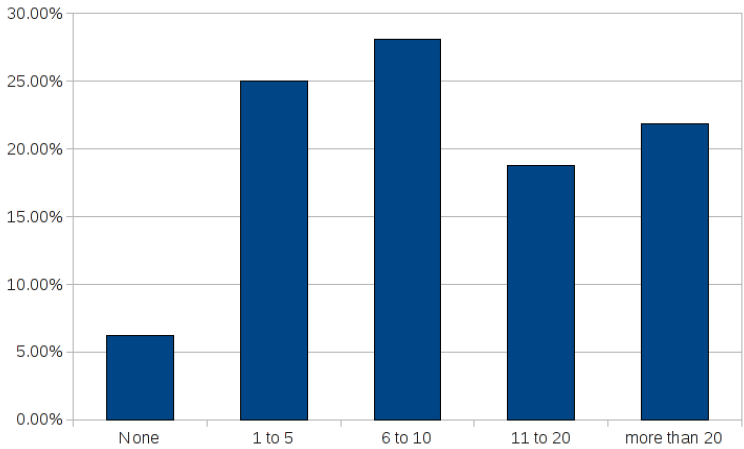

*Figure 4: Results of an anonymous survey of 32 cadets who attended an evening EI session late in the semester regarding how many EI Videos they viewed that semester. About 2/3 of video views are spread evenly through the semester with the remaining 1/3 of video views in the 48 hours before graded reviews and final exams.*

Figure 4 shows the results of an anonymous survey administered to students who attended EI one evening between the end of classes and the beginning of final exams. Only 2/32 (6%) of cadets had not watched any EI videos, showing that the videos were widely utilized among cadets attending evening EI. However, even though over 50 videos are available in the courses served, most (19/32) cadets report watching 10 or less, demonstrating that the cadets are utilizing videos as needed and not as a way to short circuit the homework process by going straight to every available video. Only 22% of respondents report watching more than 20 videos. These results agree with the total of 852 cadets in Calculus 1 viewing the videos available for that course approximately 8500 times. From a manpower point of view, providing this many in-person EI sessions would be a daunting task.

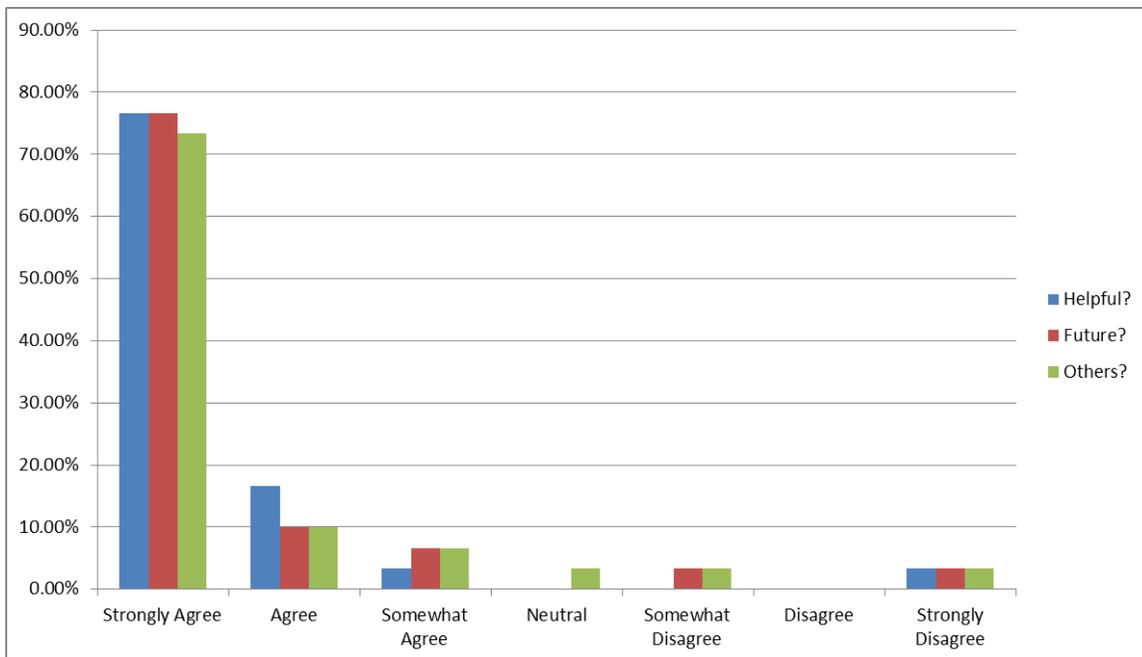

*Figure 5: I found the EI videos helpful. (blue)  I will watch EI videos in the future. (red) I will recommend EI videos to others. (green)*

Figure 5 shows the survey results of the 30 students who reported watching EI Videos. Over 70% strongly agreed that the EI videos were helpful, planned to watch EI videos in the future, and would recommend EI videos to other cadets. Less than 10% reported any negative sentiments.

**Conclusion**
Cadets and instructors also report that the demonstration of technology in EI videos also greatly increases cadet comfort with these tools and encourages their use by cadets earlier in the course. Each math course has some graphing tools that are allowed and encouraged, and often weaker students put off learning them until late in the semester.  Calculus 3 is particularly challenging for cadets who attempt to pass without learning either MATLAB or Mathematica, because these tools empower both two and three dimensional graphics, and they automate repetitive calculations with speed and accuracy that are seldom matched with pencil and paper.  EI videos demonstrating the speed, accuracy, and power of these tools in working multiple integrals, executing vector operations, computing arc length, graphing vector functions, and solving optimization problems via LaGrange multipliers have proven both popular and effective in helping cadets grow into letting the technology handle the repetitive tasks and freeing them to concentrate on interpreting the problem, choosing the appropriate tools, and applying a sound problem solving methodology without getting bogged down in the computational minutiae.  When leveraging technology in EI videos, we prefer to open with a vignette, proceed to identify the principles and develop the picture and plan on a white board, and only move to the computer at the execution stage of the solution, being deliberate to move back to the board or at least use a combination of verbal and graphical assessment before concluding.

Textbook updates and other curricular changes will probably render the first round of videos obsolete within a few years.  Given the other responsibilities of QRC faculty, keeping the videos current with the curriculum will prove challenging, and the STEM departments typically do not have the available resources to assist with video production. One idea for the future is to have cadets themselves produce EI Videos as a course project or other assignment.  We've also found that the high demand for videos can often produce slow downloading times on the USAFA intranet before graded reviews and final exams, and that the SharePoint occasionally becomes unavailable at inopportune times. We've begun hosting videos for some of the courses on outside servers and will be moving to redundant web hosting of all the videos in the near future both to ease the bottleneck on the single internal SharePoint server and to have alternate hosts when one is unavailable.

**Appendix: Production tips**

*Preparation:*
1. Work the problem in advance and plan a solution that is well presented including a vignette, big idea, picture, plan, evaluation, numerical solution, and assessment. *Components:*
    *Vignette (optional)*: A 30-120 second skit or monologue describing the relevance of the topic to the military, encouraging good practice or study habits, or generally describing the importance of Math for an officer.
    *Main Idea:* Tell them what the problem is and what the big idea will be used to solve it.
    *Picture:*  If at all possible, draw a picture that represents the problem.
    *Plan:* A plan with bullet points or numbers is a great outline for the solution you are about to present.
    *Evaluation:* This is the solving step where the plan is executed.  Don't skip steps and remind the viewer of the justification for each step.
    *Solution:* Substitute in the numbers, describing the solution, including units, etc.
    *Assessment:* Explain how you know whether the answer makes sense in the context of the original problem.  Are the units, magnitude, and sign what you expected?  Why?  Can the problem be bounded with simple estimates?
2. Plan how you will use board space to minimize erasing and poor presentation.
3. Shoot in a well-lit room and turn all lights on.
4. Plan to shoot in one take.  It is easier to edit out mistakes from one take than to splice multiple short videos into a longer one, so just back up and start again from any mistake you might make.
5. Dress nicely.  It will help your confidence.  Some trial and error is often necessary in finding clothes that look good on camera.  One instructor always wears a suit coat because his shirts never look good in back.
6. Position an academy banner or academy sports poster to the side of the board viewing area behind where you'll stand when just talking.  Subliminal message: Being at a service academy is a tremendous privilege.
7. Look through the camera and define the viewing area with some horizontal and vertical lines so that you don't write out of the viewing area while presenting your solution.   The viewing area should be completely above your waist, extend to the top of the white board or as high as you can reach and be 4-6 feet wide.  Clean the board before beginning.
8. If you wish to include a vignette, plan it ahead of time and practice it.  Keep it between 30 and 120 seconds.
9. Shoot for a video under 10 minutes after editing for Calculus 1 and 2; Calculus 3 and Physics videos often run longer.
10. Review some existing videos to get the gist of this.

*Presentation:*
1. Smile. Math is fun.  Look happy.  Be happy.  Introduce yourself with a smile, transition between segments with a smile, and learn to smile while you are pausing and thinking.  Smile when you make a mistake and when you re-enter after a mistake or pause.  Cheesy is better than sour or angry.
2. Don't stop the camera.  Splicing is much harder than editing out long pauses.
3. Be sure to present all the parts of your plan: vignette, main idea, picture, plan, evaluation, solution, and assessment.
4. Stand in front of the flag/poster when you're just talking, so you don't block the solution.
5. Don't obstruct the solution when you are pointing.  Hold your open hand, facing forward, BELOW the part of the solution you wish to highlight.
6. Write neatly, level, and large enough to be legible on the video, but not so large you need to erase a lot to remain in the viewing area.
7. Eye contact with the viewer means looking at the camera.  Instructors can turn the screen toward themselves, so checking the screen is perceived as eye contact.
8. Don't be afraid to try again.  Most folks don't produce a video they are proud of on their first try.  Copy the video to your computer, review it carefully, be intentional about the things you want to do differently, and try again on a different day.

*Editing:*
1. You need to copy the video to your hard disk for editing.  Connect the camera to your computer via USB and look for the most recent file in the directory structure.
2. For many formats, editing is easily done with Avidemux, which is freely available and can be downloaded from a number of sources.  Do a Google search and be sure you are downloading version 2.5 or later.  Avidemux runs on Windows and Linux.  The AVS video editor (for Windows)  has more features and enables titles, scrolling credits, voice overs, and better support if including different media types.  OpenShot is a free video editor for Linux with an impressive list of features, but it was extremely buggy and crash prone in our testing.
3. Avidemux is intuitive to use for cutting out unwanted material, mistakes, transitions, sneezes, etc.
4. Most single problem videos should be no longer than 10 minutes after editing.
5. Video, Audio, and Format specifications need to be carefully entered before you save the video, so that its formatting and compression are compatible with the other videos and preserve quality without being too big.
6. The specifications we use are: Video spec should be MPEG4-AVC.  Video Encoding Mode

should be "Video size, two pass."  Target size should be 4 megabytes times the number of minutes: 10 minutes -> 40 megabytes.  Other video encoding options should be left as their defaults.
   7. Audio spec should be AAC (Faac).  Other audio setting should be left as their default.
   8. Format should be mp4.
   9. These specs are a good compromise between size and quality, enabling the written solution to be readable.
   10. It takes most computers of recent vintage 2-6 times the video run time to fully compress the video.
   11. Play the video yourself to ensure all went well before uploading.

Technical note: We like the JVC Everio GZ-MG335HU video camera ("normal definition") better than the JVC Everio GZ-HD320BU video camera ("high definition").  The normal definition camera saves as JPEG files where each frame is preserved without any loss.  The high definition camera compresses the video in real time and the compression algorithm does not seem to be optimized for material written on the board.  The normal definition camera allows compression to be controlled at the editing phase.

**Acknowledgements**

The EI Video project has been funded, in part, by BTG Research (www.btgresearch.org).  Feedback and participation by Capt Justin Rufa and Capt Sean Estrada (USAFA DFMS) has also been essential to the project.